\documentclass{article}%
\usepackage{amsmath}%
\usepackage{amsfonts}%
\usepackage{amssymb}%
\usepackage{graphicx}

\begin{document}

\begin{center}
{\large On the nature of dark energy. Is the Casimir force a }

{\large manifestation of this exotic kind of energy?}\textrm{\medskip}
\end{center}

Piero\ Brovetto$^{\dag}$, Vera Maxia$^{\dag}$\ and Marcello Salis$^{\ddag}
$\ (\footnote{Corresponding Author e-mail: pbrovetto@gmail.com}

$^{\dag}$On leave from Istituto Fisica Superiore - University of Cagliari, Italy

$^{\ddag}$Dipartimento di Fisica - University of Cagliari, Italy\medskip

Summary - Topics concerning dark energy and the accelerating expansion of
universe are briefly reported. Arguments about the quantum fluctuations of
vacuum are examined in order to decide if the Casimir force really is a
laboratory evidence of dark energy.

\medskip\ 

\textbf{1 - The rate of the universe expansion}

In the last decade of the past century, intensive investigations were
undertaken in order to probe the slowing down of the universe expansion. The
most important tool was the measurement of magnitudes of type Ia supernovae
placed in remote galaxies. Since these objects are caracterized by a standard
peak brightness, this allows determination of their distances from earth and
consequently their antiquity. By measuring their Hubble red-shifts as well,
the expansion rate of universe was gauged versus time, starting from the Big
Bang event dating about fourteen billion years ago. Expansion rate was found
decreasing in the course of the first nine billions, but speeding up over the
last five billions. To account for this surprising result, cosmologists
proposed a modified form of Einstein general theory of relativity by
introducing in the field equation an additional term depending on the vacuum
energy, that is,%

\begin{equation}
G_{\mu\nu}=8\pi G\left(  T_{\mu\nu}+\rho_{VAC}\ g_{\mu\nu}\right)  ,\label{a}%
\end{equation}
where $G_{\mu\nu}$ is the space curvature tensor, $G$ the Newton gravity
constant, $T_{\mu\nu}$ the stress-energy tensor, $g_{\mu\nu}$ the metric
tensor and $\rho_{VAC}$ the vacuum energy density usually referred to as "dark
energy" $\left[  \text{1, 2}\right]  $. Term $\rho_{VAC}\ g_{\mu\nu}$,
opposite to $T_{\mu\nu}$, represents a "negative-form-of-gravity" due to the
pressure of vacuum quantum fluctuations. Since it is an intrinsic property of
space, it should remain constant even when the universe expands.

\textbf{2 - The vacuum quantum fluctuations}

Quantum fluctuations of vacuum is a topic which require some specifications.
Indeed, owing to the energy-time indeterminacy principle $\delta w\ \delta
t\simeq h$, fluctuations $\delta w=2m_{e}c^{2}$ originate in vacuum virtual
electron-positron pairs which join again after a very short time $\delta
t\simeq4\cdot10^{-21}$s. It follows that the electron-positron separation
keeps small with respect to Compton wavelength $h/m_{e}c=2.4\cdot10^{-12}$m.
Consequently, pairs act as dipoles and vacuum behaves like a polarizable
dielectric which has the effect of reducing all charges by a constant amount
$\left[  3\right]  $. In the vacuum diagrammatic picture, one electron and one
positron line starting from a vertex $A$ rejoin in a vertex $B$ while a wavy
line, representing a photon, comes back from $B$ to $A$. Six first-order
diagrams of this kind are possible, all involving one photon line. Among
these, four involve interaction with matter $\left[  4\right]  $. The
arguments just outlined show that vacuum is filled up with electromagnetic
energy. This entails the existence of an internal pressure related to energy
density, like occurs for the black-body radiation $\left[  5\right]  $.

\textbf{3 - Does the dark energy really exist?}

The previous interpretation of the accelerating rate of the universe expansion
is not free from difficulties. Indeed, all the attempts of evaluate
$\rho_{VAC}$, the dark energy density, led to absurdly large values, giving
rise to pressures so high that all matter in the universe would instantly fly
apart. Actually, the density magnitude acceptable on the ground of
cosmological arguments is somewhat small, which presumably precludes its
identification in the laboratory $\left[  1\right]  $. For this reason, a
different interpretation was recently proposed $\left[  6\right]  $. It
assumes that our galaxy, the milky way, lies in an emptier-than-average region
of space, a sort of huge void. Consequently, the reduced presence of
gravitational matter, that is, barionic and dark matter, originates a reduced
slowing down of the space expansion which could be mistaken for an
acceleration. Probably, future observations, improving the statistic bases of
these researches, will differentiate between the two interpretations.

\textbf{4 - The Casimir effect}

Owing to this disappointing state of affairs, it seems right to consider a
different argument in principle useful for understanding the nature of dark
energy. In year 1948, the Dutch physicist Hendrik Casimir proposed a peculiar
experiment for detecting the quantum fluctuations of vacuum $\left[  7\right]
$. The Casimir's experimental set-up is as follows. Two plane conducting
plates of area $A$ are positionned parallel at a small distance $d $ on the
order of one $\mu$m. In the space span between the plates, the wavelength
$\lambda$ of virtual photons propagating in direction orthogonal to the plates
must satisfy the condition $d=n\ \lambda/2$ ($n=1,2,3...$), like real photons
in a Fabry-Perot cavity. In this way, most of the propagation modes is turned
off, while in the external space all modes remain allowed. Consequently, an
imbalance in the quantum vacuum pressure follows which originates between the
plates the attractive force%

\begin{equation}
F_{C}=\frac{\pi hc}{480}\frac{A}{d^{4}},\label{b}%
\end{equation}
that is, just the Casimir force. It is expected to be rather small, indeed by
assuming $A=1\ $cm$^{2}$ and $d=1\ \mu$m, equation $\left(  2\right)  $ yields
$F_{C}=1.3\cdot10^{-7}$ N, which corresponds to pressure $F_{C}/A$ of
$1.3\cdot10^{-8}$\ b ($1\ $b$\ =0.987$ Atm). In the course of the last
half-century, theory of Casimir effect was intensively studied\ and various
experiments for measuring $F_{C}$ were carried out, but with poor success
$\left[  8\right]  $. Since this is due mainly to the difficulty of achieving
a perfect parallelism between the plates, in year 1997, a conclusive
experiment was performed utilizing a sphere-flat-plate geometry. The
experimental data were compared with a modified form of equation $(2)$ showing
an agreement of 5\% with theory $\left[  9\right]  $. Recently, utilizing the
parallel-plate configuration, equation $(2)$ was checked at the 15\% precision
level in the $0.5-3\ \mu$m range $\left[  10\right]  $.

These results are acknowledged as a direct evidence of vacuum pressure.
Consequently, by letting $P_{V}$ and $P_{P}$ be the pressures in the external
space and in space between the plates, respectively, and considering that
pressures $P_{P}$ and $F_{C}/A$ depend on plate separation $d$, we have%

\begin{equation}
P_{V}-P_{P}\left(  d\right)  =F_{C}\left(  d\right)  /A.\label{d}%
\end{equation}
Actually, when $d\rightarrow\infty$ pressure $P_{P}$ attains its maximum value
$P_{V}$. When $d$ decreases, $P_{P}$ decreases too in such a way that for
$d\rightarrow0$ we can let $P_{P}=0$. This merely because no pressure can be
accounted when space goes to zero. On this ground, by omitting contribution of
$P_{P}$ and choosing $d=1\ \mu$m, it follows from equation ($3$)
$P_{V}>1.3\cdot10^{-8}$\ b, a figure not manifestly unreliable. But, this
choice has no meaning in connection with vacuum fluctuations. It was pointed
out in Section \textbf{2 }that, owing to quickness of vacuum fluctuations, the
electron-positron separation is small with respect to Compton wavelength. By
choosing tentatively $d$ equal to this length, that is, $2.4\cdot10^{-12}\ $m,
we get $P_{V}>3.9\cdot10^{14}\ $b. This absurdly large figure is due to the
diverging character of equation ($2$) for small values of $d$. This equation
is fit for representing macroscopic phenomena, but it appears inadequate as
for vacuum fluctuations and the dark energy problem.

In our opinion and also in order to avoid possible confusion with van der
Waals force, it seems right to devise alternative experiments fit for
detecting vacuum fluctuations in laboratory. For this purpose, consider, for
istance, a light emitter placed between conducting plates, like those in
Casimir experiment. It is expected to show a modified spectrum, owing to
reduced vacuum fluctuations which, as pointed out in Section \textbf{2},
control the particle charges (\footnote{) An Interesting experiment of this
kind has been recently proposed \ based on measure of current fluctuations in
Josephson junctions $\left[  11\right]  $.}).\medskip

\textbf{References}

$\left[  1\right]  $ S. Perlmutter, Physics Today, April 2003, pag. 53
(references therein).

$\left[  2\right]  $ L.M. Krauss, M.S. Turner, Scientific American, September
2004, pag. 67.

$\left[  3\right]  $ W. Heitler "The Quantum Theory of Radiation" (Oxford at
the Clarendon Press, 1970) \S \ 32.

$\left[  4\right]  $ E.K.U. Gross, E. Runge, O. Heinonen "Many Particle
Theory" (Adam Hilger, New York, 1991) Ch. 20.

$\left[  5\right]  $ C. M\"{o}ller "The theory of relativity" (Oxford at the
Clarendon Press, 1955) \S \ 81.

$\left[  6\right]  $ T. Clifton, P.G. Ferreira, Scientific American, April
2009, pag. 32 (references therein).

$\left[  7\right]  $ H.G.B. Casimir, Proc. Kon. Ned. Akad. Wet. \textbf{51},
49 (1948).

$\left[  8\right]  $ G. Plunien, B. Muller and W. Greiner "The Casimir effect"
Phys. Report \textbf{134}, 87 (1986) (references therein).

$\left[  9\right]  $ S. K. Lamoreaux, Phys. Rev. Lett. \textbf{78}, 5 (1997).

$\left[  10\right]  $ G. Bressi, G. Carugno. R. Onofrio and G. Ruoso, Phys.
Rev. Lett. \textbf{88}, 41804 (2002).

$\left[  11\right]  $ C. Beck and M.C. Mackey, arXiv:astro-ph, 11 Dec 2006.
\end{document}